\documentclass[twocolumn,aps,prl,showpacs,amsmath,amssymb]{revtex4-1}
\usepackage{epsfig}
\usepackage{graphicx}
\usepackage{dcolumn}
\usepackage{bm}
\usepackage[colorlinks=true,dvipdfm]{hyperref}

\begin{document}
\bibliographystyle{nature}
\title{Self-similarity breaking: Anomalous nonequilibrium finite-size scaling and finite-time scaling}
\author{Weilun Yuan, Shuai Yin,}
\author{Fan Zhong}  \email{Corresponding author: stszf@mail.sysu.edu.cn}
\affiliation{State Key Laboratory of Optoelectronic Materials and Technologies, School of
Physics, Sun Yat-sen University, Guangzhou 510275, People's
Republic of China}
\date{\today}

\begin{abstract}
Symmetry breaking plays a pivotal role in modern physics. Although self-similarity is also a symmetry and appears ubiquitously in nature, a fundamental question is whether self-similarity breaking makes sense or not. Here, by identifying the most important kind of critical fluctuations dubbed as phases fluctuations and comparing the consequences of having self-similarity with those of lacking self-similarity in the phases fluctuations, we show that self-similarity can indeed be broken with significant consequences at least in nonequilibrium situations. We find that the breaking of self-similarity results in new critical exponents which give rise to violation of the well-known finite-size scaling or the less-known finite-time scaling and different leading exponents in the ordered and the disordered phases of the paradigmatic Ising model on two- or three-dimensional finite lattices when it subjects to the simplest nonequilibrium driving of linear heating or cooling through its critical point, in stark contrast to identical exponents and different amplitudes in usual critical phenomena. Our results demonstrate how surprising driven nonequilibrium critical phenomena can be. Application to other classical and quantum phase transitions is highly expected.
\end{abstract}

\maketitle
Symmetry breaking is well known and plays a pivotal role in modern physics. Although self-similarity is a kind of symmetry and appears ubiquitously in nature~\cite{Mandelbrot,Meakin}, a fundamental question is whether self-similarity breaking makes sense or not, because self-similarity holds inevitably only within a certain range in nature~\cite{Meakin}, in contrast to rigorous mathematical objects like fractal~\cite{Mandelbrot}.

Critical phenomena~\cite{Fisher,Mask,Tc} are generic systems with self-similarity up to a diverging correlation length $\xi$. The self-similarity can be limited by a finite system size $L$. Yet, it remains up to $L$ and appears as finite-size scaling (FSS)~\cite{fss,Barber,Cardy,Privman}. FSS has been verified theoretically~\cite{Brezin,BrezinZinn}, numerically~\cite{Cardy,Privman}, and even partially experimentally~\cite{Gasparini}, and is the most widely used numerical method to extract critical properties~\cite{landaubinder}. Dynamic FSS has also been confirmed~\cite{Suzuki,Wansleben}. Can FSS fail? If yes, self-similarity may be broken.

Critical phenomena are also temporally self-similar up to a diverging correlation time~\cite{Mask,Hohenberg,Folk}. To avoid the resultant critical slowing down~\cite{Swendsen,Wolff}, one can also restrict the self-similarity. This is achieved by driving a system through its critical point at a finite rate. This rate gives rise to a finite timescale that is experimentally controllable and serves as the temporal analogue of $L$ in FSS. As a result, the self-similarity is reflected in finite-time scaling (FTS)~\cite{Zhong1,Zhong2}. Moreover, the system itself is driven off equilibrium when the finite driven timescale becomes shorter than the diverging correlation time.
FTS has also been successfully applied to many systems theoretically~\cite{Yin,Yin3,Zhong2,Liu,Huang,Liupre,Liuprl,Pelissetto,Xu,Xue,Feng,Cao,Gerster,Li,Mathey} and experimentally~\cite{Clark,Keesling}. Its renormalization-group theory~\cite{Zhong06} has been generalized to the case with a weak driving of an arbitrary form, leading to a series of driven nonequilibrium critical phenomena such as negative susceptibility and competition of various regimes and their crossovers, as well as violation of fluctuation-dissipation theorem and hysteresis~\cite{Feng}. Again, violation of FTS may signal breaking of self-similarity.

When both spatial and temporal limitations are present~\cite{Zhong2}, a special revised FTS containing both $L$ and the rate of cooling was suggested and verified exactly at the critical point~\cite{Liu,Huang}. This gives rise to a distinctive leading exponent for the order parameter $M$ in cooling~\cite{Huang}. However, no violation of FSS and FTS was discovered. What then about the whole driving process rather than just at the critical point?

Here, by linearly heating and cooling the paradigmatic Ising model whose equilibrium critical properties are well known, we show that spatial and temporal self-similarities of the most important part of critical fluctuations that is dubbed as phases fluctuations (PsFs)---the plural form of phase here is both to emphasize that at least two phases are involved owing to symmetry breaking and to distinguish it from the usual phase of a complex field---can indeed be broken in a series of driven nonequilibrium critical phenomena in which the simple FSS and FTS for some observables are violated in both heating and cooling. Moreover, the breaking of self-similarity, or bressy in short, results in bressy exponents that give rise to different leading exponents for the ordered and disordered phases both upon heating and upon cooling, in stark contrast to identical exponents but different amplitudes in equilibrium critical phenomena. Further, the bressy exponents are probably new critical exponents under heating despite combinations of known critical exponents under cooling. Although questions such as how bressy results in the bressy exponents and how different behaviors cross over are yet to be studied, the results found here demonstrate how surprising driven nonequilibrium critical phenomena can be. Application to other classical and quantum phase transitions is thus highly expected.

We first recapitulate the theories of FSS and FTS.
Consider a system of a size $L$ driven from one phase through a critical point at $T_c$ to another phase by changing the temperature $T$ with a finite rate $R>0$ such that
\begin{equation}
T-T_c\equiv\tau=\pm Rt,
\label{rate}
\end{equation}
where $+$ ($-$) corresponds to heating (cooling). We have chosen $t=0$ at $T_c$ for simplicity.
We start with the scaling hypothesis for the susceptibility $\chi$
\begin{equation}
\chi(\tau, R, L^{-1})=b^{\gamma/\nu}\chi(\tau b^{1/\nu}, Rb^{r},L^{-1}b),
\label{RG}
\end{equation}
which can be derived from the renormalization-group theory~\cite{Brezin,BrezinZinn,Zhong06,Zhong1,Zhong2}, where $b$ is a scaling factor, $\gamma$, $\beta$, $\nu$, and $r$ are the critical exponents for $\chi$, $M$, $\xi$, and $R$, respectively. We have replaced the time $t$ with $R$ because they are related by Eq.~(\ref{rate}). Moreover, from the same equation, we find~\cite{Zhong}
$r=z+1/\nu$
because $t$ transforms as $tb^{-z}$ with $z$ being the dynamic critical exponent~\cite{Mask,Hohenberg,Folk}.

From Eq.~(\ref{RG}), choosing the length scale $b = R^{-1/r}$ leads to the FTS form~\cite{Zhong06,Zhong1,Zhong2,Huang}
\begin{equation}
\chi=R^{-\gamma/r\nu}\mathcal{F}_{T\chi} (\tau R^{-1/r\nu},L^{-1} R^{-1/r}),
\label{FTSX}
\end{equation}
while assuming $b =L$ results in
\begin{equation}
\chi=L^{\gamma/\nu}\mathcal{F}_{S\chi}(\tau L^{1/\nu},RL^r),
\label{FSSX}
\end{equation}
which is the FSS form under driving, where $\mathcal{F}_{T\chi}$ and $\mathcal{F}_{S\chi}$ are universal scaling function. Similarly, $M$ must behave~\cite{Zhong06,Zhong1,Zhong2,Huang}
\begin{equation}
M=R^{\beta/r\nu}F_{TM} (\tau R^{-1/r\nu},L^{-1} R^{-1/r})
\label{FTSM}
\end{equation}
in the FTS regime, while in the FSS regime, it becomes
\begin{equation}
M=L^{-\beta/\nu}F_{SM}(\tau L^{1/\nu},RL^r),
\label{FSSM}
\end{equation}
where $F_{TM}$ and $F_{SM}$ are also scaling functions.

Usually, the scaling functions are analytic for vanishingly small scaling variables~\cite{Zhong1,Zhong2,Huang}. This implies $R^{-1/r}\ll|\tau|^{-\nu}$ and $R^{-1/r}\ll L$ in the FTS regime, for example. In other words, the driven length scale $R^{-1/r}$ is the shortest among $\xi\sim |\tau|^{-\nu}$ and $L$. Therefore, in the FTS (FSS) regime, $L^{-1}R^{-1/r}$ ($RL^r$) is negligible and the leading singularity is just the factor in front of each scaling function. If $L^{-1}R^{-1/r}$ ($RL^r$) is large, crossover to FSS (FTS) regime occurs.

In equilibrium critical phenomena, one can define different critical exponents above and below $T_c$~\cite{Fisher,Mask,Tc}. However, they are identical because of the absence of singularities across the critical isotherm~\cite{Fisher}. Only the amplitudes of the leading singularities and thus the scaling functions above and below $T_c$ differ. However, we will see distinctions upon driving in the following.

Consider the standard Ising model with the Hamiltonian
$\mathcal{H}=-J\sum_{\langle i,j\rangle} \sigma_{i}\sigma_{j}$
which describes the interaction $J>0$ of a spin $\sigma_i=\pm 1$ on site $i$ of a simple square or cubic lattice with its nearest neighbors. Periodic boundary conditions are applied throughout. We employ the single-spin Metropolis algorithm~\cite{MC} and interpreted it as dynamics~\cite{Glauber,landaubinder}. We prepared the system in ordered or disordered initial configurations and then heated and cooled it, respectively, according to a given $R$. We checked that the initial states generate no differences once they are sufficiently far away from $T_c$. $30~000$ samples were used to average. Tripling that number only smoothes the curves without appreciable displacements.  We study mainly the two-dimensional (2D) model whose $T_c=2J/\ln(1+\sqrt{2})\approx2.269$, $\beta=1/8$, $\nu=1$, $\gamma=7/4$~\cite{Mask}, and $z=2.167$~\cite{z2d}. For the 3D model, $T_c =J/0.221~659~5(26)=4.511~42(6)$~\cite{exponent2}, $\nu = 0.630~1(4)$, $\beta = 0.326~5(3)$, $\gamma =1.237~2(5)$~\cite{Tc,exponent1,exponent2}, and $z=2.055$~\cite{Huang,z3d1,z3d2}.

We study the following observables
\begin{equation}
\left\langle m\right\rangle= \left\langle\frac{1}{L^d}\sum_{i=1}^{L^d} \sigma_{i}\right\rangle,~\left\langle |m|\right\rangle =\left\langle\left|\frac{1}{L^d}\sum_{i=1}^{L^d} \sigma_{i}\right|\right\rangle,
\label{M}
\end{equation}
\vskip -0.4cm
\begin{equation}
\chi=L^d\left(\left\langle m^2\right\rangle-{\left\langle m\right\rangle}^2\right), ~\chi'=L^d\left(\left\langle m^2\right\rangle-{\left\langle |m|\right\rangle}^2\right),
\label{modelII}
\end{equation}
where the angle brackets represent ensemble averages and $d$ is the space dimensionality. We will generally refer to $M$ and $\chi$ for both definitions and specify to a specific one when so indicated.

\begin{figure}
  \centerline{\epsfig{file=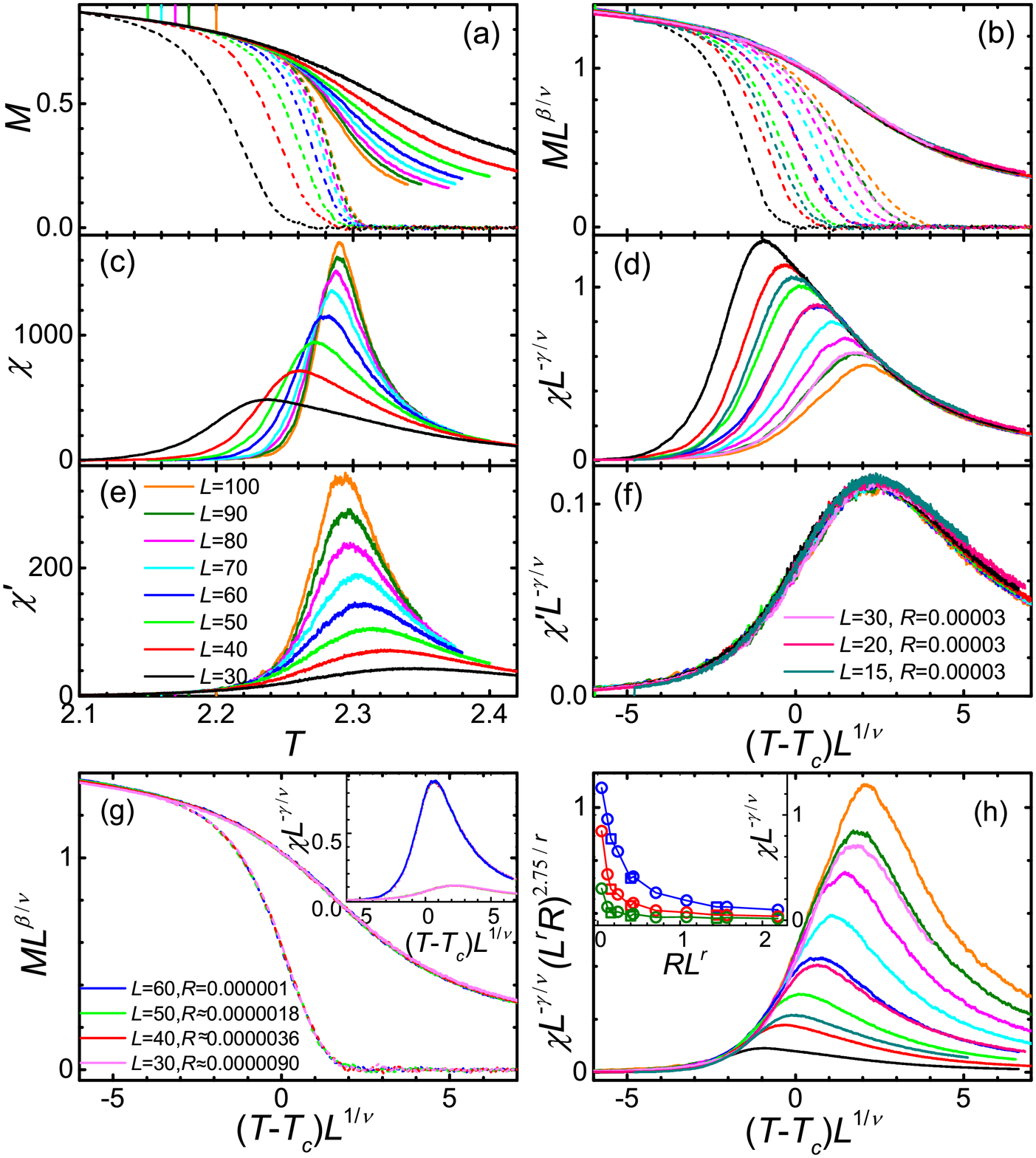,width=1.0\columnwidth}}
  \caption{\label{fssh} (Color online) FSS in heating of the 2D Ising model at fixed rates. (a) $M$ and (b) its rescaled form and (c)--(f), $\chi$ and its rescaled form at $R=0.000~001$ on different lattice sizes given in the legend in (e). (g) FSS of $M$ and $\chi$ (inset) for $RL^r=10^{-6}\times 60^{3.167}\approx0.433$ on four lattice sizes given in the legend in (g). (h), revised FSS for the ordered phase. The inset in (h) displays the dependence on $RL^r$ of $\chi L^{-\gamma/\nu}$ at $\tau L^{1/\nu}=-1$ (blue), $-2$ (red), and $-3$ (olive) for $R=0.000~03$ (squares) and $0.000~001$ (circles). The errorbar of each datum is estimated to be not larger than the symbol size. Lines connecting symbols are only a guide to the eye. In (a), (b) and (g), the dashed lines represent results of $\langle m\rangle$ and $\chi$ and the solid lines those of $\langle |m|\rangle$ and $\chi'$. In (b), (d), (f), and (h), curves of another fixed $R=0.000~03$ on three different lattices given in the legend in (f) are also shown.}
\end{figure}
We show in Fig.~\ref{fssh} $M$ and $\chi$ in heating for fixed rates. It is prominent that $\langle |m|\rangle$ and $\langle m\rangle$ differ and the peak temperatures of $\chi$ and $\chi'$ even exhibit a qualitatively different dependence on $R$. However, what is the most remarkable is that the FSS of both $\langle m\rangle$ and $\chi$ at the low-$T$ side---referred to roughly as the ordered phase in the following---are completely violated, though that of $\langle |m|\rangle$ and $\chi'$ and the high-$T$ side or the disordered phase of $\chi$ are good. In fact, it is $\langle |m|\rangle$ and $\chi'$ that were utilized in the early verification of FSS~\cite{Landau76,landaubinder}.

The stark difference between $\langle m\rangle$ and $\langle |m|\rangle$ originates from PsFs. It indicates that $m$ can overturn and fluctuate from the spin-up phase to the spin-down phase. These flips between the two phases appear more often and finally on a par as $T$ increases, resulting in the vanishing $\langle m\rangle$ at high temperatures. Moreover, $\langle |m|\rangle$ and $\langle m\rangle$ of different lattice sizes converge to an envelop at low temperatures, implying that both phases assume the equilibrium magnetization. These qualify us to call the fluctuating clusters of predominantly up or down spins and of the size $\xi$ as phases. These PsFs result in the large critical fluctuations, which is apparent from the large difference between $\chi$ and $\chi'$ in Figs.~\ref{fssh}(c) and~\ref{fssh}(e). This is because critical fluctuations originate directly from the symmetry-broken phases.

The violation of FSS stems from the bressy in the PsFs. Note first that corrections to scaling~\cite{Wegner} are only slight even for $L=15$ as can be appreciated from the high-$T$ side in Figs.~\ref{fssh}(d) and~\ref{fssh}(f). Note also that once $RL^r$ is fixed, the FSS recovers completely as Fig.~\ref{fssh}(g) demonstrates. However, this does not mean that we have to consider both scaled variables in Eqs.~(\ref{FSSM}) and~(\ref{FSSX}) in a 3D space~\cite{Cao}. The good scalings for various $R$ values of $\langle |m|\rangle$, $\chi'$, and $\chi$ in the disordered phase indicate that $RL^r$ can be safely ignored and hence the system can be considered quasi-equilibrium. Yet, the $\langle m\rangle$ curves in Fig.~\ref{fssh}(a) can only occur in nonequilibrium. Otherwise, those parts that are different from $\langle |m|\rangle$ must be averaged to zero owing to the PsFs. Therefore, the fixed $RL^r$ serves, instead, to ensure the same survival time of the fluctuating phases for different $R$ and therefore the temporal self-similarity of the PsFs, because it is equivalent to a fixed ratio of the average time for a cluster to overturn, $L^z$, and the driven time $R^{-z/r}$ during which the driving changes appreciably~\cite{Yin,Huang}.

We now find a bressy exponent defined as an extra singularity originates from the bressy. Since the dependence differs in the ordered and the disordered phases, we distinguish them by the subscripts $\mp$, respectively, though a single $\mathcal{F}_{S\chi'}$ and $F_{S\langle |m|\rangle}$ are enough as seen in Figs.~\ref{fssh}(b) and~\ref{fssh}(f). From the inset in Fig.~\ref{fssh}(h), the dependence of $\mathcal{F}_{S\chi-}$ on $RL^r$ is clearly singular. Indeed, the good collapse in Fig.~\ref{fssh}(h) shows that $\mathcal{F}_{S\chi-}(\tau L^{1/\nu},RL^r)\propto(RL^r)^{\sigma/r}$ with $\sigma=-2.75\pm0.15$ in consistence with the power-law exponent fitted out from the inset in Fig.~\ref{fssh}(h). We note, however, that for large $RL^r$, the collapse gets poor; while for $R\rightarrow0$, the standard FSS ought to recover and a crossover may occur. Nevertheless, with $\sigma$, the leading behavior of $\chi$ in heating is $L^{\gamma/\nu+\sigma}R^{\sigma/r}$ in the ordered phase, manifestly different from the usual $L^{\gamma/\nu}$ in the disordered phase, in sharp contrast to just the amplitude difference in equilibrium critical phenomena.

For the 2D Ising model, $-\sigma$ may be $(2\gamma-6\beta)/\nu$ or $(\gamma+8\beta)/\nu$ or other combinations such as $d+6\beta/\nu$ using $2\beta+\gamma=d\nu$~\cite{Mask,Fisher}. Yet, these expressions are strange and thus $\sigma$ is more likely a new exponent. However, in the 3D model, we find $\sigma=-d\pm0.15$ simply~\cite{Yuan}, markedly different from any possible 2D expression, though their numerical values slightly overlap. Theories and results from other models are thus highly in need.

The same $\sigma$ accounts for the violated $\langle m\rangle$ scaling as well and corroborates the bressy mechanism. Indeed, from Eqs.~(\ref{modelII}),~(\ref{FSSX}), and~(\ref{FSSM}), $F_{S\langle m\rangle-}^2\propto\mathcal{F}_{S\chi'}+F_{S \langle |m|\rangle}^2-(RL^r)^{\sigma/r}$ singularly, though the first two terms on the right-hand side are analytic because of the analyticity of $\chi'$ and $\langle |m|\rangle$. In the following, we will not consider the observable with such regular terms since bressy always results in a pair like $\chi$ and $\langle m\rangle$.

\begin{figure}
  \centerline{\epsfig{file=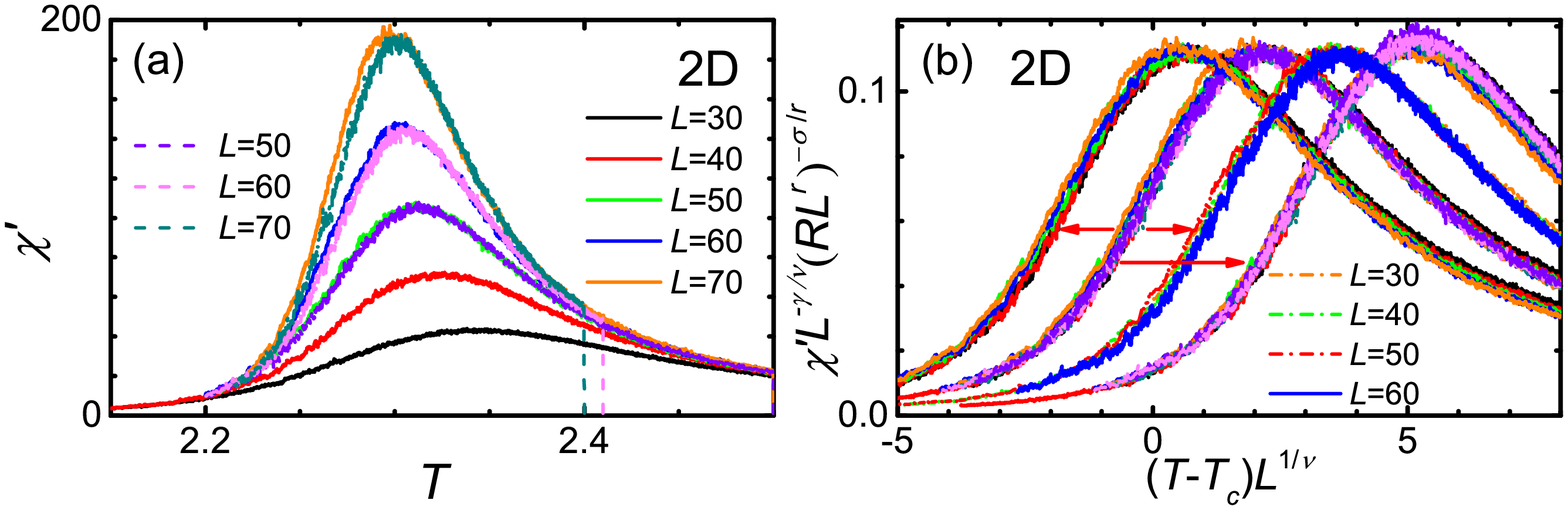,width=1.0\columnwidth}}
  \vskip -0.1cm
  \centerline{\epsfig{file=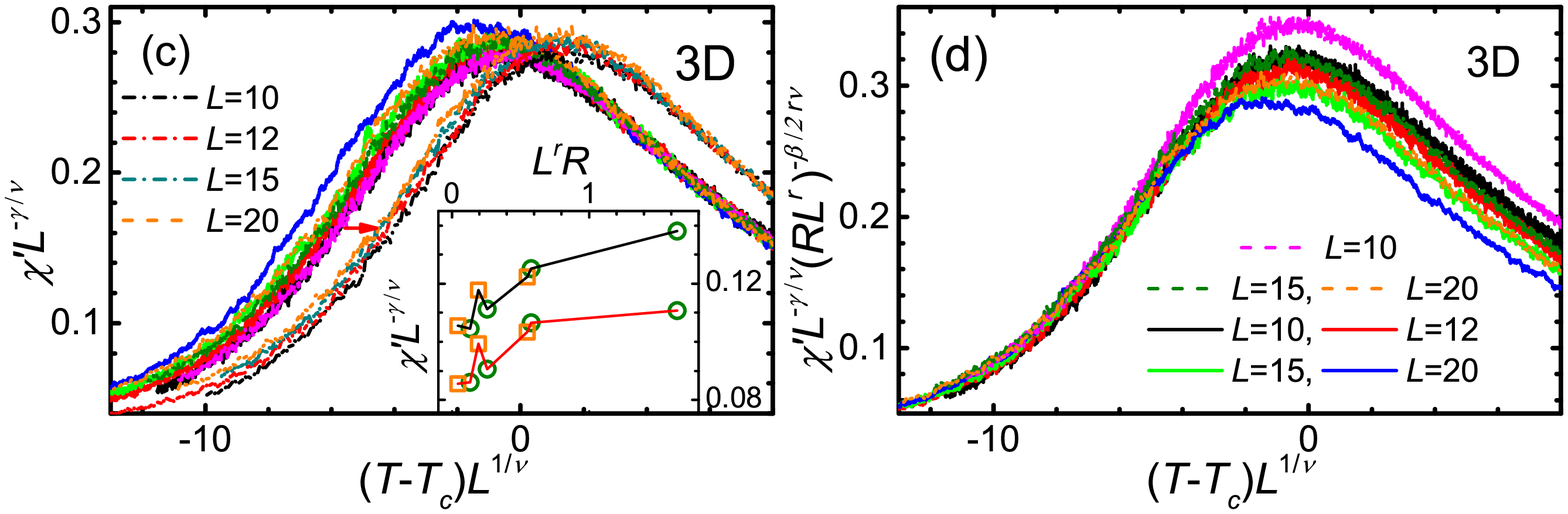,width=1.0\columnwidth}}
  \caption{\label{fssc} (Color online) (a) $\chi'$ and (b)--(d) the FSS of the 2D, (a) and (b), and the 3D, (c) and (d), Ising models in cooling with fixed $R=0.000~003$ (solid lines) and $0.000~000~3$ (dashed lines) in (a) and (b), and $R=0.000~03$ (solid lines) and $R=0.000~01$ (dashed lines) in (b) and (d). The arrows indicate the displacements of the curves by $\mp1.5$ and $3$ in (b) and $2$ in (c) for clarity. In (b), the rightmost curve has $\sigma=\beta/2\nu$ while the other three curves have $\sigma=0$. The leftmost curve includes only $R=0.000~003$. The curves shifted by $1.5$ in (b) and $2$ in (c) have fixed $RL^{r}\approx1.30$ and $0.548$, respectively, determined by the largest $L$ of the four lattices used given in the legends in them. The inset in (c) depicts $\chi'L^{-\gamma/\nu}$ at $\tau L^{1/\nu}=-8$ (black) and $-9$ (red) vs $RL^r$ for $R=0.000~01$ (squares) and $0.000~03$ (circles). The errors can be as large as double the symbol sizes due to the small $R$ and the lines are only a guide to the eye.}
\end{figure}
Under cooling, $\langle m\rangle$ is vanishingly small due to the absence of a symmetry-breaking field. $\langle m\rangle$ and $\langle |m|\rangle$ are thus completely different, indicating again the role of the PsFs. Now, the FSS of all the observables studied appears good both for one or several $R$ values in the 2D model at first sight, as seen in the first two curves on the left in Fig.~\ref{fssc}(b) for $\chi'$. However, with a $\sigma=\beta/2\nu$, the collapse in the ordered phase appears better and that in the disordered phase worse, as the rightmost curve in Fig.~\ref{fssc}(b) manifests itself, though $\chi$ exhibits no such behavior. This again demonstrates possible different leading singularities in the two phases.

In the 3D model, as seen in Fig.~\ref{fssc}(c), the systematic dependence of $\chi'$ on $R$ for the two $R$ values in the ordered phase is more visible. The dependence on $RL^r$ appears singular and, from Fig.~\ref{fssc}(d), a $\sigma=\beta/2\nu$ again collapses quite well the ordered phase and is consistent with the power-law exponent from the inset in Fig.~\ref{fssc}(c). Note that in cooling, the 2D and 3D expressions of $\sigma$ are identical, different from heating. The large 3D $\sigma$ value should be responsible for the visibility.

\begin{figure}[b]
\centering
\centerline{\epsfig{file=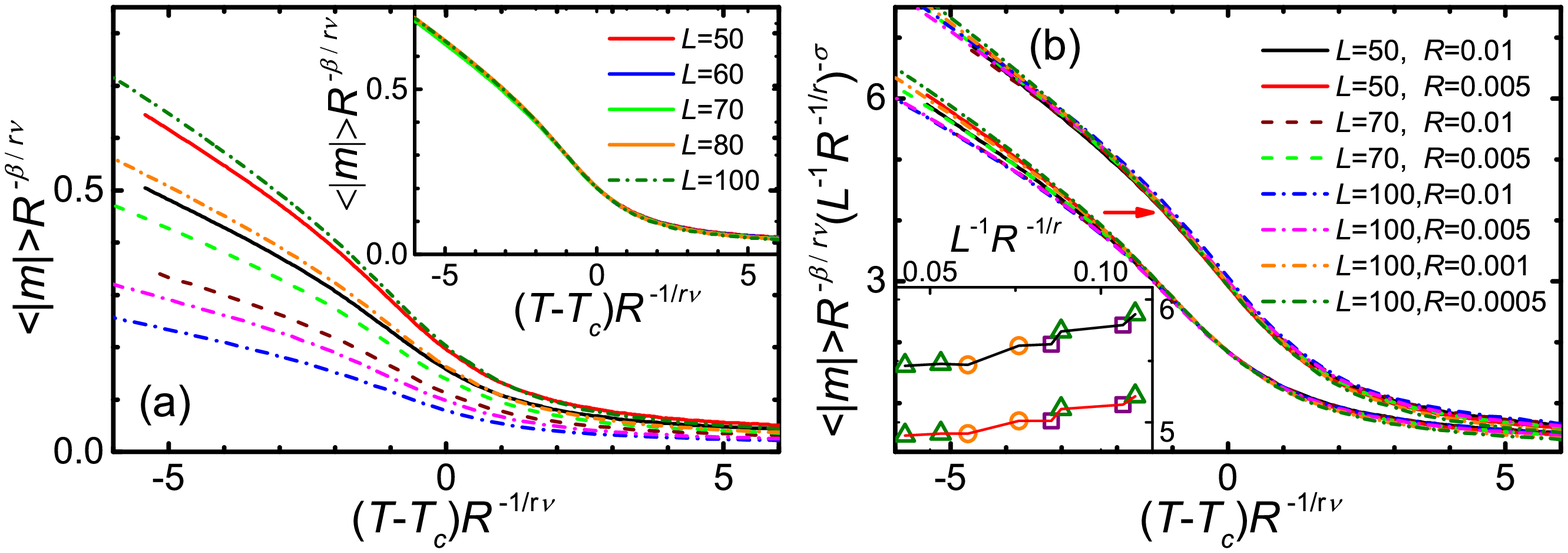,width=1.0\columnwidth}}
\caption{(Color online) (a) FTS of $\langle|m|\rangle$ and (b) its revised form in cooling on three fixed 2D lattices given in the legend in (b). In (a), the inset depicts the FTS of $\langle|m|\rangle$ for fixed $L^{-1}R^{-1/r}=100^{-1}\times0.0005^{-1/3.17}\approx0.110$ on the lattices indicated. In (b), the left curve has $\sigma=d/2$ while the right curve has $\sigma=d/2+\beta/2\nu$ and is shifted by $0.8$ as the arrow indicates. Besides, the inset displays $\langle|m|\rangle R^{-\beta/r\nu}(L^{-1}R^{-1/r})^{-d/2}$ at $\tau R^{-1/r\nu}=-4$ (red) and $-5$ (black) for $L=50$ (squares), $70$ (circles), and $100$ (triangles). Errorbars and Lines are similar to those in Fig.~\ref{fssh}.}
\label{ftsc}
\end{figure}
The violation of FSS in cooling appear not as clear cut as those in heating. The collapses with a fixed $RL^r$ from Figs.~\ref{fssc}(b) and~\ref{fssc}(c) also behave differently in the two phases, and the 3D case even shows a slightly systematic dependence on $L$. Yet, this must stem from corrections to scaling, especially for the small $L$. Nonetheless, the differences between having and lacking self-similarity as seen in Figs.~\ref{fssc}(a) and~\ref{fssc}(b) as well as~\ref{fssc}(c) are evident, especially for the latter, and
consistent with the finite $\sigma$.

In Fig.~\ref{ftsc}(a), we display a complete violation of FTS for $\langle |m|\rangle$ in cooling, though the FTS of $\chi$ and even $\chi'$ are reasonably good. Because $\langle m\rangle$ is vanishingly small and different from $\langle |m|\rangle$, the PsFs are pivotal. Indeed, FTS becomes almost perfect once lattices of different sizes hold identical number of the phases of size $R^{-1/r}$ and thus the spatial self-similarity~\cite{Li} of the PsFs is ensured by fixing $L^{-1}R^{-1/r}$. Moreover, the fluctuating phases must satisfy the central limit theorem and thus behave as $L^{-d/2}$ for large $L$~\cite{Huang}. This implies that $F_{T\langle |m|\rangle}\propto (L^{-1}R^{-1/r})^{d/2}$ singularly~\cite{Huang}, which is confirmed by the left curve in Fig.~\ref{ftsc}(b). Such a singularity has been invoked to rectify the leading behavior and results in the distinctive leading exponent in cooling exactly at $T_c$~\cite{Huang}. However, no violation of FSS and FTS was discovered. Further, the ordered phase seems weakly singular as the inset in Fig.~\ref{ftsc}(b) shows. Indeed, a further $\beta/2\nu$ consistent with the power-law exponent in the inset renders the collapse in the ordered phase better albeit worse in the disordered phase, demonstrating again the different leading singularities in the two phases. Moreover, similar to FSS in cooling, the same $\sigma$ applies to the 3D model as well~\cite{Yuan}.

\begin{figure}
  \centerline{\epsfig{file=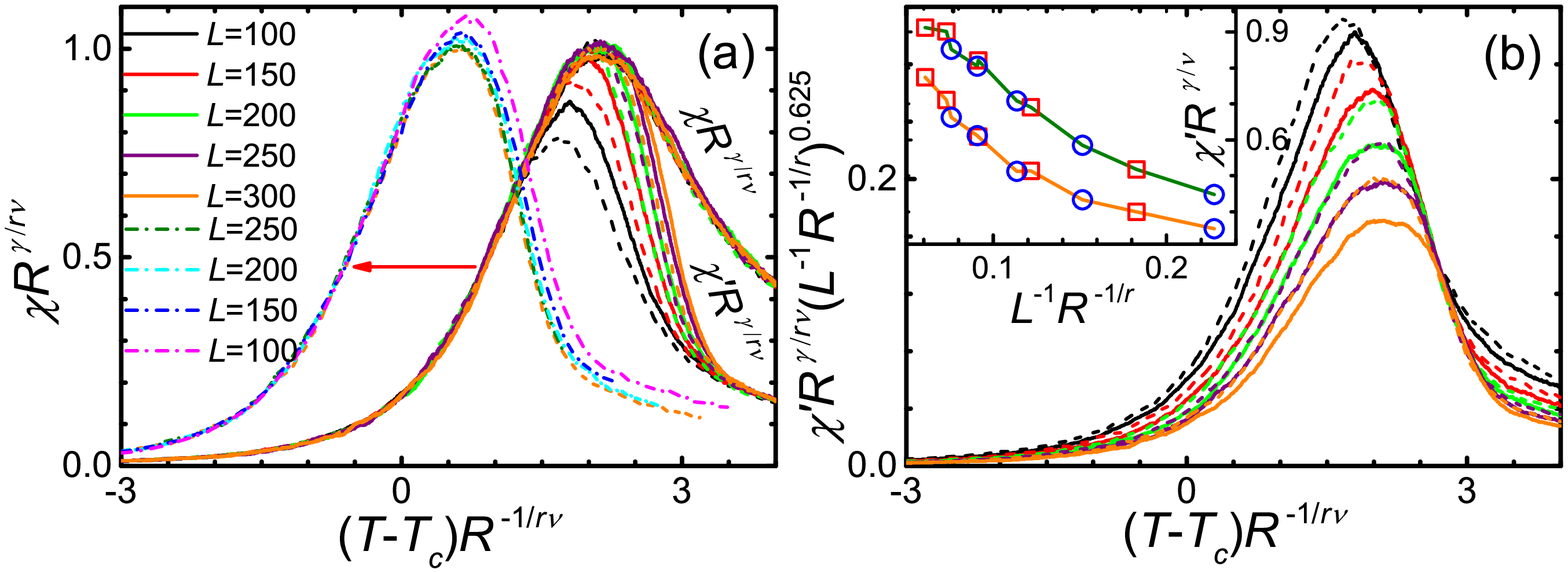,width=1.0\columnwidth}}
  \caption{\label{ftsh} (Color online) FTS in heating of (a) $\chi$ and $\chi'$, marked on the right, and (b) the revised FTS of $\chi'$ on fixed 2D lattice sizes given by the first five items in the legend. The left curve in (a) shifted by $-1.5$ indicated by the arrow has fixed $L^{-1}R^{-1/r}=300^{-1}\times0.000~05^{-1/r}\approx0.0758$ containing the last five lattice sizes in the legend. Solid and dashed lines of identical colors have $R=0.000~1$ and $0.000~05$, respectively. The inset in (b) displays the dependence on $L^{-1}R^{-1/r}$ of $\chi' R^{\gamma/r\nu}$ at $\tau R^{-1/r\nu}=2.5$ (olive) and $2.7$ (orange) for $R=0.000~1$ (squares) and $0.000~05$ (circles). Errorbars and Lines are similar to those in Fig.~\ref{fssh}.}
\end{figure}
In Fig.~\ref{ftsh}(a), we shows the FTS of $\chi$ in heating on several fixed lattices. It is apparent that $\chi$ and $\chi'$ differ and thus PsFs are again indispensable. Different from FSS, it is the FTS of $\chi'$ in the disordered phase that is violated, while $\chi$ and $\langle m\rangle$ display good FTS for the $R$ employed and require only a single scaling function. Having and lacking self-similarity apparently differs from the two sets of curves with fixed and non-fixed $L^{-1}R^{-1/r}$, respectively. Full self-similarity recovers as the curves get closer to each other for larger $L$ and smaller $R$ due to decreasing corrections to scaling. Again, the dependence on $L^{-1}R^{-1/r}$ is singular, as illustrated in the inset in Fig.~\ref{ftsh}(b). Indeed, choosing $\sigma=-0.625\pm0.025$ in consistence with the power-law exponent in the inset collapses the curves in the disordered phases, showing again the different leading behaviors in the two phases.
Here, $\sigma$ can be $(4\beta-\gamma)/2\nu$ and others, but again, is more likely a new exponent. It is definitely different from $-2\beta/\nu$ to $(2\beta-\gamma)/\nu$, which are $-0.6530/\nu$ to $-0.5842/\nu$ found for the 3D model~\cite{Yuan}, though the numerical values given are again overlapped.

\begin{acknowledgments}
This work was supported by the National Natural Science Foundation of China (Grant No. 11575297).
\end{acknowledgments}



\end{document}